\documentclass[12pt]{article}
\usepackage[cp1251]{inputenc}
\usepackage[english,russian,ukrainian]{babel}

\begin{document}

\begin{center}
\begin{large}
\textbf{On the Fermi -- Bose duality of the Dirac equation with nonzero mass}
\end{large}
\vskip 0.5cm

\textbf{V.M. Simulik and I.Yu. Krivsky}
\end{center}

\begin{center}
\textit{Institute of Electron Physics, National Academy of
Sciences of Ukraine, 21 Universitetska Str., 88000 Uzhgorod,
Ukraine}
\end{center}

\begin{center}
\textit{E-mail: sim@iep.uzhgorod.ua}
\end{center}

\vskip 1.cm

\noindent We have proved on the basis of the symmetry analysis of
the standard Dirac equation with nonzero mass that this equation
may describe not only fermions of spin 1/2, but also bosons of
spin 1. The new bosonic symmetries of the Dirac equation in both
the Foldy -- Wouthuysen and the Pauli -- Dirac representations
are found. Among these symmetries (together with the 32-dimensional pure
matrix algebra of invariance) the new, physically meaningful spin
1 Poincare symmetry of equation under consideration is proved. In order to carry out the
corresponding proofs a 64-dimensional extended real Clifford --
Dirac algebra is put into consideration.

\vskip 0.5cm


\vskip 0.5cm


\vskip 1.cm

\section{Introduction}

The main physically meaningful result of this paper is the
following. It has been shown that the Dirac equation (with nonzero
mass!) can describe equally well both spin 1/2 fermionic field
and spin 1 bosonic field. The Dirac's well-known result
(complemented by Foldy -- Wouthuysen (FW) analysis and the
Bargman -- Wigner classification of fields), which is presented
in each handbook on quantum mechanics, contains only one
(fermionic) half of our consideration.

An interest in the problem of the relationship between the Dirac
and Maxwell equations, which describe a spin 1 field, emerged
immediately after the creation of quantum mechanics [1-11]. The
bosonic spin 1 representation of the Poincare group $\mathcal{P}$,
the vector (1/2,1/2) and the tensor-scalar $(1,0) \otimes (0,0)$
representations of the Lorentz group $\mathcal{L}$, as the groups
of invariance of the \textit{massless} Dirac equation, were found
in our papers [12-17]. Here $\mathcal{P}$ is the universal
covering of the \textit{proper ortochronous} Poincare group
$\mbox{P}_ + ^ \uparrow = \mbox{T(4)}\times )\mbox{L}_ + ^
\uparrow $ and $\mathcal{L}$ = SL(2,C) is that of the Lorentz group
$\mbox{L}_ + ^ \uparrow $ = SO(1,3), respectively. In this paper, using the conception of \textit{the relativistic invariance} of the theory, we consider only its invariance with respect to some
representations of the proper ortochronous Poincare group (the
group $\mathcal{P}$ without reflections, CPT, etc.)

This paper is directly related to our previous papers [12-17]. It
is the next logical step, which follows from considerations
[12-17], where each symmetry of the massless Dirac equation is
mapped into the corresponding symmetry of the Maxwell equations
in terms of field strengths (generated by the gradient-type
current), and vice versa. Let us remind that
$\mathcal{P}$-symmetries of a field equation (its invariance with
respect to a certain representation of the group $\mathcal{P}$) have
the principal physical meaning: the Bargman -- Wigner
classification of the fields is based on the
$\mathcal{P}$-symmetries of the corresponding field equations.
Relationship between the massless Dirac equation and
above mentioned Maxwell equations [12-17] means, in particular, that each
equation is invariant not only with respect to the fermionic
$\mathcal{P}$-representation (for the $s$=1/2 spin doublet), but
also with respect to the bosonic tensor-scalar
$\mathcal{P}$-representation (with $s$=1,0).

Now we are able to present the similar results for the general
case when the mass in the Dirac equation is nonzero.

First we seek for the additional symmetries of the FW equation
[18]. Than one can easily find the corresponding symmetries of the
standard Dirac equation fulfilling the well-known FW
transformation [18]. Thus, we find a wide set of symmetries of
the Dirac equation in both FW and standard Pauli -- Dirac (PD)
representations. The 32-dimensional algebra $\mbox{ A}_{32}$ of
invariance is found. In the FW representation, the algebra
$\mbox{ A}_{32}$ is the maximal algebra of invariance in the
class of \textit{pure matrix operators} (without any derivatives
over the space variables). The physically meaningful spin 1
symmetries are found here as subalgebras of the $\mbox{ A}_{32}$
algebra. Finally, we have found the new hidden Poincare symmetry
of the Dirac equation and fulfilled the Bargman -- Wigner
analysis of corresponding $\mathcal{P}$-representation. This
analysis proved that this hidden symmetry is spin 1 Poincare
symmetry of the Dirac equation. On the basis of last assertions
we come to a conclusion that the Dirac equation may also describe
the spin 1 field, i.e. not only fermions of spin 1/2, but also
bosons of spin 1. Thus, below the investigation of the spin 1
symmetries of the Dirac equation with nonzero mass started in [19, 20] is being continued.

In Sec. 2, a brief review of the general assertions about Poincare
invariance and covariance of the spinor field theory is given.
The ordinary handbook consideration is rewritten in terms of
so-called \textit{primary} generators, which are associated with the
real (not imaginary) parameters of the groups of invariance. Few
tiny and hidden assumptions of group-theoretical approach to the
spinor field theory are explained.

In Sec. 3, in the canonical FW-representation, the extended real
Clifford -- Dirac algebra (ERCD) is put into consideration.

In Sec. 4, the 32-dimensional maximal pure matrix algebra $\mbox{ A}_{32}$ of
invariance of the FW equation is found.

In Sec. 5, the spin 1 Lorentz-symmetries of the FW and Dirac equations
are obtained.

In Sec. 6, the spin 1 Poincare-symmetry of the FW equation is proved.

In Sec. 7, the spin 1 Poincare-symmetry of the standard Dirac equation
is put into consideration.

In Sec. 8, the brief remix of our previous results [12--17] for
bosonic symmetries of the massless Dirac equation is presented.
The specific characters of the case $m=0$ in comparison with
nonzero mass are considered briefly. The possibility, which is
open here, of generalization of all results [12--17] to the
general case of nonzero mass is mentioned.

In Sec. 9, the brief general conclusions are given.

We use the units $\hbar=c=1$; the summation over a twice repeated index is performed.

\section{Notations, assumptions and definitions}

Here we consider the Dirac equation

\begin{equation}
\label{eq1} \left( {i\gamma ^\mu \partial _\mu - m} \right)\psi
\left( x \right) = 0; \quad \partial_{\mu}\equiv\partial/\partial x^{\mu},\mbox{ }\mu = \overline {0,3} =
0,1,2,3,\mbox{ }\psi \in \mathrm{S}(\mathrm{M}(1,3))\times \mathrm{C}^{4}\equiv\mathrm{S}^{4,4},
\end{equation}

\noindent as an equation in the \textit{test function space}
$\mathrm{S}^{4,4}$ of the 4-component functions over the Minkowski
space M(1,3). We note that the complete set $\{\psi\} \equiv\Psi$
of solutions of the equation (1) contains the generalized
solutions belonging to the space
$\mathrm{S}^{4,4*}\supset\mathrm{S}^{4,4}$ of the Schwartz's
\textit{generalized} functions, $\{\psi^{\mathrm{gen}}\} \subset
\mathrm{S}^{4,4*}$. The mathematically well-defined consideration
of this fact demands (see, e.g., the book [21] on the
axiomatic approach to the field theory) the functional representation
of the elements of $\psi^{\mathrm{gen}} \subset
\mathrm{S}^{4,4*}$, which makes the consideration very
complicated. Hence, we recall that the \textit{test function
space} $\mathrm{S}^{4,4}$ is \textit{dense} in
$\mathrm{S}^{4,4*}$. It means that any element
$\psi^{\mathrm{gen}} \subset \mathrm{S}^{4,4*}$ can be
approximated (with arbitrary degree of accuracy) by an element
$\psi \in \mathrm{S}^{4,4}$ from the corresponding Cauchy
sequence in the space $\mathrm{S}^{4,4}$. Therefore, here for the
equation (1) we have restricted ourselves to the supposition
$\psi \in \mathrm{S}^{4,4}$. Such supposition is physically
verified and essentially simplifies consideration without
any loss of generality and mathematical correctness.

The invariance of equation (1) with respect to the hidden
transformations will be considered. We seek for
the Bose, not Fermi, symmetries. In order to fulfill this
programme let us briefly recall the conventional relativistic
invariance of equation (1). In the modern consideration, it
is the invariance with respect to the universal covering
$\mathcal{P}\supset \mathcal{L}$=SL(2,C) of the proper
ortochronous inhomogeneous Poincare group $\mbox{P}_ + ^ \uparrow
\supset \mbox{L}_ + ^ \uparrow $ = SO(1,3). For our purposes it is
convenient to rewrite in the space $\mathrm{S}^{4,4}$ the
standard Fermi (spinor) representation of the group $\mathcal{P}$
in terms of the \textit{primary} $\mathcal{P}$-generators
$(p\equiv(p_{\mu}), \, j\equiv(j_{\mu\nu}))$. By the definition,
these generators are associated with the \textit{real parameters}
$(a, \, \omega)\equiv (a^{\mu}, \,
\omega^{\mu\nu}=-\omega^{\nu\mu})$ with well-known physical meaning. The generators $(p_{\mu}, \, j_{\mu\nu})$ are the
orts of the \textit{real} $\mathcal{P}$-\textit{algebra}, i. e.
the algebra over the real numbers (we use the similar symbols for
groups and their algebras). The necessity of using the primary
$\mathcal{P}$-generators and the real $\mathcal{P}$-algebra will
be evident below in our search for the additional (hidden)
symmetries of the equation (1).

In arbitrary representations, the primary $\mathcal{P}$-generators
$(p_{\mu}, \, j_{\mu\nu})$, as the orts of the real Lie
$\mathcal{P}$-algebra, satisfy the commutation relations

\begin{equation}
\label{eq2} [p_\mu,p_\nu ] = 0,\mbox{ }[p_\mu,j_{\rho\sigma}] =
g_{\mu\rho}p_\sigma-g_{\mu\sigma}p_\rho,\mbox{
}[j_{\mu\nu},j_{\rho\sigma}] = - g_{\mu\rho}j_{\nu\sigma} -
g_{\rho\nu}j_{\sigma\mu} - g_{\nu\sigma}j_{\mu\rho} -
g_{\sigma\mu}j_{\rho\nu},
\end{equation}

\noindent and generate a $\mathcal{P}$-representation, which is
defined by an exponential series

\begin{equation}
\label{eq3}(a, \, \omega)\in \mathcal{P} \rightarrow \widehat{\mathrm{F}}(a, \, \omega) = \exp (a^{\mu}p_{\mu}+\frac{1}{2} \omega^{\mu\nu} j_{\mu\nu}) \stackrel {\mathrm{inf}}{=} 1+a^{\mu}p_{\mu}+\frac{1}{2} \omega^{\mu\nu} j_{\mu\nu},
\end{equation}

\noindent where symbol $\stackrel {\mathrm{inf}}{=}$ defines
"infinitesimally, i. e. in the neighborhood of the unit element of the
group $\mathcal{P}$".

The primary generators $(p_{\mu}, \, j_{\mu\nu})$ for the
ordinary \textit{local} spinor (Fermi)
$\mathcal{P}$-representation in $\mathrm{S}^{4,4}$ are the
following \textit{Lie operators}

\begin{equation}
\label{eq4} p_{\rho}=\partial_{\rho}\equiv \partial /\partial x^{\rho}, \quad  j_{\rho\sigma}=m_{\rho\sigma}+s_{\rho\sigma} \quad  (m_{\rho\sigma}\equiv x_{\rho} \partial_{\sigma}-x_{\sigma} \partial_{\rho}, \quad s_{\rho\sigma} \equiv \frac{1}{4}[\gamma_{\rho},\gamma_{\sigma}]).
\end{equation}

\noindent The operators $p_{\mu}, \, j_{\mu\nu}$ from (4) commute
with the Diracian $i\gamma ^\mu \partial _\mu - m$. Therefore,
formulae (3) with generators (4) define the
\textit{local} spinor (Fermi)
$\mathcal{P}^{\mathrm{F}}$-representation of the group
$\mathcal{P}$ in the form

\begin{equation}
\label{eq5} \psi (x) \rightarrow \psi^{\prime} (x) = [\widehat{\mathrm{F}}(a, \, \omega) \equiv\widehat{\mathrm{F}}_{1}(\omega) \widehat{\mathrm{F}}_{2}(a, \, \omega)] \psi (x) \stackrel {\mathrm{inf}}{=} (1+a^{\mu}p_{\mu}+\frac{1}{2} \omega^{\mu\nu} j_{\mu\nu}) \psi (x),
\end{equation}

\noindent  which is the $\mathcal{P}$-\textit{group of invariance} of equation (1). In formula (5) the following notations are used:

\begin{equation}
\label{eq6} \psi, \, \psi^{\prime} \in  \mathrm{S}^{4,4}, \quad
\widehat{\mathrm{F}}_{1}(\omega)\equiv
\exp(\frac{1}{2}\omega^{\mu\nu}s_{\mu\nu}) \stackrel
{\mathrm{inf}}{=} (1+\frac{1}{2} \omega^{\mu\nu}s_{\mu\nu}) \sim
(\frac{1}{2},0)\otimes(0,\frac{1}{2}),
\end{equation}

\begin{equation}
\label{eq7} \widehat{\mathrm{F}}_{2}(a, \, \omega) \psi (x) \equiv \exp(a^{\mu}p_{\mu}+\frac{1}{2}\omega^{\mu\nu}m_{\mu\nu})\psi (x) = \psi (\Lambda ^{-1}(x-a)) \stackrel {\mathrm{inf}}{=} (1+a^{\mu}p_{\mu}+\frac{1}{2} \omega^{\mu\nu} m_{\mu\nu}) \psi (x), \quad \Lambda ^{-1}\in \mathrm{P}_ + ^ \uparrow.
\end{equation}

We pay attention to the following detail of the mathematical
correctness of consideration. The space $\mathrm{S}^{4,4}$ is
the common domain of definitions and values both for the
generators $(p_{\mu}, \, j_{\mu\nu})$ (4) and for all functions
from them, which we use here (in particular, for the exponential
series (3) convergent in the space $\mathrm{S}^{4,4}$).
Further, we mark that usually the fermionic
$\mathcal{P}$-transformations of the field $\psi$ are written  in
the form

 \begin{equation}
\label{eq8} \psi (x) \rightarrow \psi^{\prime} (x^{\prime}) = \widehat{\mathrm{F}}_{1}(\omega) \psi (x),\quad x^{\prime}=\Lambda x+a,\quad \Lambda \in \mathrm{P}_ + ^ \uparrow.
\end{equation}

\noindent However, this form (contrary to formula (5))
does not demonstrate manifestly the mathematical definition of
the group of invariance of equation (1), which is given by
the definition: $\mathcal{P}^{\mathrm{F}}$ is a group of
invariance of equation (1), if  $\mathcal{P}^{\mathrm{F}}\Psi
= \Psi$; or for arbitrary solution of equation (1): if from
$\psi(x)\in\Psi$ results $\widehat{\mathrm{F}}(a, \, \omega) \psi (x)\in
\Psi \subset \mathrm{S}^{4,4}$, where $\widehat{\mathrm{F}}(a, \, \omega)$ is given by (5).

Let us note further that in conventional field-theoretical approach, instead of our primary Lie generators (4), the following operators are used

\begin{equation}
\label{eq9} p^{\mathrm{stand}}_{\rho}=ip_{\rho}\equiv i\partial_{\rho}, \quad
j^{\mathrm{stand}}_{\rho\sigma}=ij_{\rho\sigma}\equiv m^{\mathrm{stand}}_{\rho\sigma}+s^{\mathrm{stand}}_{\rho\sigma}
\quad  (m^{\mathrm{stand}}_{\rho\sigma}\equiv ix_{\rho}
\partial_{\sigma}-ix_{\sigma} \partial_{\rho}, \quad
s^{\mathrm{stand}}_{\rho\sigma} \equiv
\frac{i}{4}[\gamma_{\rho},\gamma_{\sigma}]).
\end{equation}

\noindent It is evident that these generators are associated with the pure imaginary parameters $(-ia^{\mu}, \, -i\omega^{\mu\nu})$ and with corresponding complex $\mathcal{P}$-algebra. Below, as well as in other our publications on the symmetries, primary generators (4) and customary operators (9) should not be confused.

We recall that both pure matrix operators $s_{\rho\sigma}$ and
pure differential operators $m_{\rho\sigma}$ from (4) satisfy the
same commutation relations as $[j_{\mu\nu},j_{\rho\sigma}]$ in
(2). However, the operators $s_{\rho\sigma}$ and $m_{\rho\sigma}$
(contrary to their sum $j_{\rho\sigma}$) \textit{are not}
the symmetry operators (operators of invariance) of equation
(1) (operator $\widehat{q}$ is called \textit{a symmetry
operator} or the operator of invariance of equation (1), if
equality $\widehat{q} \Psi = \Psi$ is valid, where $\Psi
\equiv \{\psi\} \subset \mathrm{S}^{4,4}$ is the complete set of
solutions of equation (1), see e. g. the corresponding definition in [22]). Therefore, both pure matrix
$\mathcal{L}$-representation $\widehat{\mathrm{F}}_{1}(\omega)$
(6) and infinite-dimensional $\mathcal{L}$-representation

\begin{equation}
\label{eq10} \widehat{\mathrm{F}}_{2}(\omega) \psi (x) \equiv
\exp(\frac{1}{2}\omega^{\mu\nu}m_{\mu\nu})\psi (x) = \psi
(\Lambda ^{-1}(x)) \stackrel {\mathrm{inf}}{=} (1+\frac{1}{2}
\omega^{\mu\nu} m_{\mu\nu}) \psi (x)
\end{equation}

\noindent in $\mathrm{S}^{4,4}$ are not the groups of invariance of equation (1). As a consequence of these facts (see arbitrary conventional consideration in relevant papers, handbooks and monographs), both matrix $\mathcal{L}$-spin $s_{\rho\sigma}$ and orbital angular momentum $m_{\rho\sigma}$ (differential  $\mathcal{L}$ angular momentum) do not generate the conserved in time integral constants of motion, i. e. being taken separately both spin and orbital angular momenta of the field $\psi$ are not conserved. Therefore, both pure matrix $\widehat{\mathrm{F}}_{1}(\omega)$ (6) and infinite-dimensional $\widehat{\mathrm{F}}_{2}(\omega)$ representations in $\mathrm{S}^{4,4}$ of the Lorentz group $\mathcal{L}$ are not the groups of invariance of the Dirac equation (1).

Besides the local $\mathcal{P}^{\mathrm{F}}$-representation, the
so-called \textit{induced} (see, e. g. [23, 24])
$\mathcal{P}^{\mathrm{F}}$-representation for the field $\psi$ is
useful and meaningful. Mathematically this representation can be
related to the \textit{conception of special role of the time
variable} $t\in (-\infty, \infty)\subset$M(1,3). Indeed, in the
general consideration of the Dirac equation in the axiomatic
approach, it follows from equation (1) that the Dirac field
$\psi$ satisfies identically the Klein -- Gordon equation

\begin{equation}
\label{eq11}(\partial^{\mu}\partial_{\mu}+m^{2} \equiv
\partial_{0}^{2}-\triangle + m^{2})\psi (x)=0, \quad \psi \in
\mathrm{S}^{4,4*},
\end{equation}

\noindent which is the equation of hyperbolic type. As a
consequence of this fact, the generalized solutions of the Dirac
equation (1) are the \textit{ordinary functions} of the time
variable $x^{0}=t\in (-\infty, \infty)\subset$M(1,3) (they are the
generalized functions of the variables
$\overrightarrow{x}\equiv (x^{\ell})\in \mathrm{R}^{3}\subset
\mathrm{M}(1,3)$ only). Therefore, due to a special role of the time
variable $x^{0}=t\in (x^{\mu})$ (in obvious analogy with
nonrelativistic theory), one can use in general consideration
\textit{the quantum-mechanical rigged Hilbert space}

\begin{equation}
\label{eq12}\mathrm{S}^{3,4}\subset \mathrm{H}^{3,4} \subset
\mathrm{S}^{3,4*};\quad \mathrm{S}^{3,4}\equiv
\mathrm{S}(\mathrm{R}^{3})\times \mathrm{C}^{4},
\end{equation}

\noindent where

\begin{equation}
\label{eq13}\mathrm{H}^{3,4}\equiv
\mathrm{L}_{2}(\mathrm{R}^{3})\times \mathrm{C}^{4} \equiv \{f:
\mathrm{R}^{3}\rightarrow \mathrm{C}^{4}, \, \int
d^{3}x|f(t,\overrightarrow{x})|^{2} <\infty\},
\end{equation}

\noindent is the quantum-mechanical Hilbert space of the 4-component
functions over $\mathrm{R}^{3}\subset \mathrm{M}(1,3)$ (depending parametrically on $x^{0}=t$), which are the square
modulus integrable over \textit{the Lebesgue measure} $d^{3}x$ in
the space $\mathrm{R}^{3}\subset \mathrm{M}(1,3)$. Just the
space $\mathrm{R}^{3}$ is interpreted as the coordinate spectrum
of the quantum-mechanical particles described by the
field $\psi$.

In this concept, the Dirac equation in the
Schr$\mathrm{\ddot{o}}$dinger form

\begin{equation}
\label{eq14} i\partial_{0}\psi = H\psi\leftrightarrow (\partial_{0}-\widetilde{p}_{0})\psi=0;\quad H\equiv
\gamma^{0}(-i\gamma^{k}\partial_{k}+m), \quad \widetilde{p}_{0}=-iH,
\end{equation}

\noindent (which is completely equivalent to the equation (1)) in
the integral form is given by

\begin{equation}
\label{eq15} \psi(t)= u(t_{0},t)\psi(t_{0});\quad u(t_{0},t)=\exp
[-iH(t-t_{0})];\quad \psi(t_{0}),\psi(t)\subset \mathrm{S}^{3,4*}.
\end{equation}

\noindent In formula (15), the unitary  operator $u(t_{0},t)$
(with arbitrary-fixed parameters $t_{0},t\in
(-\infty,\infty)\subset \mathrm{M}(1,3)$) is the operator of
automorphism in the rigged Hilbert space (12) (below we set
$t_{0}=0$).

Recall that the test function space $\mathrm{S}^{3,4}$ has
a few wonderful features. It consists of the functions being
infinitely smooth  (infinitely differentiable with respect to
$x^{\ell}$) rapidly decreasing at the infinity
$|\overrightarrow{x}|\rightarrow \infty$ in arbitrary direction
in $\mathrm{R}^{3}$ together with its derivatives of arbitrary
orders. Further, it contains the space of \textit{the finite functions}
with the same properties. Moreover, the space $\mathrm{S}^{3,4}$ is \textit{kernel} in the
triple (12). The last one means that this space is dense both in
the $\mathrm{H}^{3,4}$ and $\mathrm{S}^{3,4*}$ spaces. Therefore,
below we restrict our consideration to the suggestion
$\psi\in\mathrm{S}^{3,4}$ in equation (14). Such restriction is
both mathematically correct and technically appropriate (it does
not require the using of the functional form of the elements
$\psi\in\mathrm{S}^{3,4*}$). It is also physically motivated.
Indeed, an arbitrary measurement of a construction from
$\psi^{\mathrm{gen}} \subset \mathrm{S}^{3,4*}$ by an equipment of
an arbitrary degree of accuracy can be successfully approximated
(with forward fixing arbitrarily precise degree of accuracy) by the
corresponding constructions from the prelimit functions
$\psi\in\mathrm{S}^{3,4}\subset\mathrm{H}^{3,4}\subset\mathrm{S}^{3,4*}$.

For definiteness below we use the Pauli -- Dirac (PD) representation of the Clifford -- Dirac $\gamma$-matrices

\begin{equation}
\label{eq16} \gamma ^\mu:\mbox{ }\gamma ^\mu \gamma ^\nu + \gamma
^\nu \gamma ^\mu = 2g^{\mu \nu },\mbox{ }g = (g_\nu ^\mu ) =
\mathrm{diag}\mbox{ }g( + - - \mbox{ } - ),
\end{equation}

\begin{equation}
\label{eq17} \gamma ^0 = \left| {{\begin{array}{*{20}c}
 1 \hfill & 0 \hfill \\
 0 \hfill & { - 1} \hfill \\
\end{array} }} \right|,\mbox{ }\gamma ^k = \left| {{\begin{array}{*{20}c}
 0 \hfill & {\sigma ^k} \hfill \\
 { - \sigma ^k} \hfill & 0 \hfill \\
\end{array} }}\right|;\quad \sigma ^1 = \left| {{\begin{array}{*{20}c}
 0 \hfill & 1 \hfill \\
 1 \hfill & 0 \hfill \\
\end{array} }} \right|,\mbox{ }\sigma ^2 = \left| {{\begin{array}{*{20}c}
 0 \hfill & { - i} \hfill \\
 i \hfill & 0 \hfill \\
\end{array} }} \right|,\mbox{ }\sigma ^3 = \left| {{\begin{array}{*{20}c}
 1 \hfill & 0 \hfill \\
 0 \hfill & { - 1} \hfill \\
\end{array} }} \right|\mbox{; }\ell = 1,2,3.
\end{equation}

\noindent In this representation, the SU(2)-spin primary generators
are given by the matrices $s_{\ell n} \equiv
\frac{1}{4}[\gamma_{\ell},\gamma_{n}]$. Two of them are
quazidiagonal and the $s_{z}$-matrix is diagonal:

\begin{equation}
\label{eq18} \overrightarrow{s}=
(s_{23},s_{31},s_{12})\equiv(s^{\ell})= \frac{1}{2}\left|
{{\begin{array}{*{20}c}
 \overrightarrow{\sigma} \hfill & 0 \hfill \\
 0 \hfill &  \overrightarrow{\sigma} \hfill \\
\end{array} }} \right|\rightarrow s_{z}\equiv s^{3}=\frac{1}{2}\left|
\begin{array}{cccc}
 1 & 0 & 0 & 0\\
 0 & -1 & 0 & 0\\
0 & 0 & 1 & 0\\
0 & 0 & 0 & -1\\
\end{array} \right|.
\end{equation}

\noindent Therefore, the operator $s_{z}$ contains the projections
of the spin $\overrightarrow{s}$ on the axis z of the
quantum-mechanical spin-$\frac{1}{2}$ doublet of particles.

The primary $\mathcal{P}$-generators of the
$\mathcal{P}$-representation in the space $\mathrm{S}^{3,4}$ have
the form of the following anti-Hermitian operators

\begin{equation}
\label{eq19} \widetilde{p}_0 = -iH\equiv
-\gamma_{0}(\gamma^{\ell}\partial_{l}+ im), \mbox{
}\widetilde{p}_{k}=p_{k}= \partial _k ,\mbox{
}\widetilde{j}_{kl}=j_{kl} = m_{kl}+s_{kl}, \mbox{
}\widetilde{j}_{ok} = t\partial _k - \frac{1}{2}\{x_{k},\widetilde{p}_0\},
\end{equation}

\noindent where
$\{\mathrm{A},\mathrm{B}\}\equiv\mathrm{A}\mathrm{B}+\mathrm{B}\mathrm{A}$
and $t\in(-\infty,\infty)$ is the arbitrary-fixed parameter (the primary generators $\widetilde{q}=(\widetilde{p}_{\mu},\,\widetilde{j}_{\mu\nu})$ (19) coincide with the corresponding operators  $-i\widetilde{q}^{\mathrm{stand}}$, where $\widetilde{q}^{\mathrm{stand}}$ are given by the formulae (126)--(129) in [24]). Using the Heisenberg commutation relations in the form
$[x^{\ell},\partial_{j}]=\delta_{\ell j}$ and the SU(2)-relations
for $s^{\ell}$ (18), it is easy to show that the operators (19)
satisfy the commutation relations of $\mathcal{P}$-algebra in the
manifestly covariant form (2). Furthermore, generators (19)
commute with the operator $(\partial_{0}-\widetilde{p}_0)$ of the
equation (14). Therefore, as a consequence of anti-Hermiticity of
arbitrary operator from (19), the induced
$\mathcal{P}$-representation

\begin{equation}
\label{eq20}(a, \, \omega)\in \mathcal{P} \rightarrow
\widetilde{\mathrm{U}}(a, \, \omega) = \exp
(a^{\mu}\widetilde{p}_{\mu}+\frac{1}{2} \omega^{\mu\nu}
\widetilde{j}_{\mu\nu})
\end{equation}

\noindent in the space $\mathrm{S}^{3,4}$ is unitary and is the
group of invariance of equation (14). Hence, in the definition

\begin{equation}
\label{eq21}w^{\mu}\equiv
\frac{1}{2}\varepsilon^{\mu\nu\rho\sigma}\widetilde{p}_{\rho}\widetilde{j}_{\nu\sigma}
\rightarrow w^{0}=s^{\ell}\partial_{\ell}\equiv
\overrightarrow{s}^{\mathrm{stand}}\cdot
\overrightarrow{p}^{\mathrm{stand}},
\end{equation}

\noindent the main Casimir operators for the generators (18) are
given by

\begin{equation}
\label{eq22}\widetilde{p}^{\mu}\widetilde{p}_{\mu}\equiv
\widetilde{p}_{0}^{2}-\partial_{\ell}^{2}\equiv
m^{2}\mathrm{I}_{4}, \quad \mathrm{I}_{4}\equiv
\mathrm{diag}(1,1,1,1),
\end{equation}

\begin{equation}
\label{eq23}\mathcal{W}\equiv w^{\mu}w_\mu =
m^{2}\overrightarrow{s}^{2}=m^{2}\frac{1}{2}(\frac{1}{2} +
1)\mathrm{I}_{4}.
\end{equation}

\noindent According to the Bargman -- Wigner classification,
just this fact means that in equation (14) (for which the
induced $\mathcal{P}^{\mathrm{F}}$-representation (20) is the
group of invariance) the field $\psi$ is the Fermi field (the
field of quantummechanical spin-$\frac{1}{2} $ doublet of
particles with the mass $m$).

The relativistic invariance of the spinor field theory with
respect to the representation (20) has some special features.  It
should be stressed once more that in the induced
$\mathcal{P}^{\mathrm{F}}$-representation (20) the time $t=x^{0}$
plays a special role in comparison to the role of space
variables $x^{\ell}$. Moreover, we use in the definition (13) such
$\mathcal{P}$-non-covariant objects as the Lebesgue measures
$d^{3}x$ (or $d^{3}k$ in the momentum representation of the rigged
Hilbert space (12) for the field states
$\psi\in\mathrm{S}^{3,4}$). Nevertheless, the theory of the
spinor field $\psi$ based on the induced
$\mathcal{P}^{\mathrm{F}}$-representation (20) is obviously
relativistic invariant. The proof is given in the text after formulae
(19) and by the Bargman -- Wigner analysis of the Casimir operators
(22), (23).

Now we are in position to give some comparison of the local and
induced $\mathcal{P}^{\mathrm{F}}$-representations (5) and (20).
It is easy to see that in the set of solutions $\Psi=\{\psi\}$ of
equations (14)=(1) the local $\mathcal{P}$-representation (5)
and the induced $\mathcal{P}$-representation (20) coincide.
Moreover, the main Casimir operators for the Lie
$\mathcal{P}^{\mathrm{F}}$-generators (4) have the form

\begin{equation}
\label{eq24} p^{\mathrm{loc}\;2}\equiv
p^{\mu}p_{\mu}\equiv\partial^{\mu}\partial_{\mu}, \quad
\mathcal{W}^{\mathrm{loc}}\equiv m^{2}\overrightarrow{s}^{2}=\frac{3}{4}\partial^{\mu}\partial_{\mu}.
\end{equation}

\noindent Therefore, the eigenvalues of the main Casimir's for the
$\mathcal{P}^{\mathrm{F}}$-representations (5) and (20) coincide.
Note, by the way, that as it is easy to show formulae (24) have the same form for
arbitrary Lie operators (4) (i. e., for arbitrary local
$\mathcal{P}$-representations), for which the Lorentz spins
$s_{\mu\nu}$ generate the $\mathcal{L}$-representations
$(s,0)\otimes(0,s)$.

Mark that the local generators (4) in $\mathrm{S}^{4,4}$
are the functions of 14 independent operators
$x^{\rho},\,\partial_{\rho},\,s_{\rho\sigma}$ (the Lorentz spin
operators $s_{\rho\sigma}$ are the independent orts of the
CD-algebra). The conception of Hermiticity or anti-Hermiticity in
$\mathrm{S}^{4,4}$ is not inherent both for these 14 operators and
$\mathcal{P}^{\mathrm{F}}$-generators (4). Therefore, the
concept of unitarity is not inherent for the local
$\mathcal{P}$-representation (5) as well. It means that the
$\mathcal{P}$-representation (5) itself (and, similarly, its generators
(4)) does not contain the information "what quantum-mechanical
particles are described by the filed $\psi$ from equation (1)".

Contrary to these facts, the generators (19) of induced
$\mathcal{P}^{\mathrm{F}}$-representation in the space
$\mathrm{S}^{3,4}$ are the anti-Hermitian functions of
(particularly) other 11 independent operators
$\overrightarrow{x}=(x^{\ell}),\,\partial_{\ell},\,s_{k\ell},\,\gamma^{0},\,m$
or are the functions of the standard Hermitian in
$\mathrm{S}^{3,4}$ operators: $\overrightarrow{x}=(x^{\ell}),\,
\overrightarrow{p}^{\mathrm{stand}}=(-i\partial_{\ell}), \,
\overrightarrow{s}^{\mathrm{stand}}=(is_{\ell n }),\,
\gamma^{0},\,m$.
Hence, the induced $\mathcal{P}^{\mathrm{F}}$-representation (20)
is unitary in the quantum-mechanical rigged Hilbert space (12).
However, the restriction of the local
$\mathcal{P}^{\mathrm{F}}$-representation (5) on the set
$\Psi\subset\mathrm{S}^{3,4}$ of solutions of equation
(1)=(14) coincides with the induced
$\mathcal{P}^{\mathrm{F}}$-representation (20).

Nevertheless, both local and induced
$\mathcal{P}^{\mathrm{F}}$-representations have few common
physical shortcomings (namely it is the reason of our start from the
FW-representation of the spinor field theory, where
$\mathcal{P}^{\mathrm{F}}$-representation does not have the above
shortcomings). These shortcomings are related to the PD-representation of the
spinor field $\psi$. In order to explain these assertions, we
recall the general solution of equation (14)=(1) in the
rigged Hilbert space (12):

\begin{equation}
\label{eq25} \psi(x)=\frac{1}{(2\pi)^{\frac{3}{2}}}\int
d^{3}k[a^{\mathrm{r}}(\overrightarrow{k})v_{\mathrm{r}}^{-}(\overrightarrow{k})e^{-ikx}+b^{*\mathrm{r}}(\overrightarrow{k})v_{\mathrm{r}}^{+}(\overrightarrow{k})e^{ikx}],\quad
\mathrm{r}=1,2.
\end{equation}

\noindent Here $kx\equiv \widetilde{\omega} t -
\overrightarrow{k}\overrightarrow{x}, \quad \widetilde{\omega}\equiv\sqrt
{\overrightarrow{k}^{2} + m^2}$  and the 4-component spinors are
given by

\begin{equation}
\label{eq26} v_{\mathrm{r}}^{-}(\overrightarrow{k})=N\left|
{{\begin{array}{*{20}c}
 (\widetilde{\omega}+m)\mathrm{d}_{\mathrm{r}}\\
 (\overrightarrow{\sigma}\cdot \overrightarrow{k})\mathrm{d}_{\mathrm{r}}\\
\end{array} }} \right|,\quad v_{\mathrm{r}}^{+}(\overrightarrow{k})=N\left|
{{\begin{array}{*{20}c}
 (\overrightarrow{\sigma}\cdot \overrightarrow{k})\mathrm{d}_{\mathrm{r}}\\
 (\widetilde{\omega}+m)\mathrm{d}_{\mathrm{r}}\\
\end{array} }} \right|;\quad
N\equiv\frac{1}{\sqrt{2\widetilde{\omega}(\widetilde{\omega}+m)}},\quad
\mathrm{d}_{1}=\left| {{\begin{array}{*{20}c}
 1\\
 0\\
\end{array} }} \right|,\quad \mathrm{d}_{2}=\left|
{{\begin{array}{*{20}c}
 0\\
 1\\
\end{array} }} \right|.
\end{equation}

\noindent We want to emphasize that, in general, the four independent
amplitudes
$a^{\mathrm{r}}(\overrightarrow{k}),\,b^{\mathrm{r}}(\overrightarrow{k})$
are the generalized functions from the rigged Hilbert space in the
momentum representation. Still, if $\psi \in
\mathrm{S}^{3,4}\subset\mathrm{S}^{3,4*}$, then
$a^{\mathrm{r}}(\overrightarrow{k}),\,b^{\mathrm{r}}(\overrightarrow{k})\in
\mathrm{\widetilde{S}}^{3,4} \equiv
\mathrm{\widetilde{S}}(\mathrm{R}_{\vec{k}}^{3})\times\mathrm{C}^{4}$
(where $\mathrm{R}_{\vec{k}}^{3}$ is the spectrum of the
quantummechanical operator $\overrightarrow{p}$) and the space
$\mathrm{\widetilde{S}}^{3,4}$ is a kernel in the rigged Hilbert
space
$\mathrm{\widetilde{S}}^{3,4}\subset\mathrm{L}_{2}(\mathrm{R}_{\vec{k}}^{3})\times\mathrm{C}^{4}
\subset\mathrm{\widetilde{S}}^{3,4*}$. Taking into account that
any partner from the triplet of spaces (12) is invariant with
respect to the 3-dimensional Fourier-transformation, one can omit the
symbol "tilde" over these spaces. Therefore, just the
fundamental solutions $e^{\mp
ikx}v_{\mathrm{r}}^{\mp}(\overrightarrow{k})$ of the equations
(14)=(1) are the essentially generalized solutions (for them
$(a^{\mathrm{r}}(\overrightarrow{k}),\,b^{\mathrm{r}}(\overrightarrow{k}))\sim\delta(\overrightarrow{k}-\overrightarrow{k}^{\prime})\in\mathrm{S}(\mathrm{R}_{\vec{k}}^{3})^{*}$).

It is well-known that the spinors
$v_{\mathrm{r}}^{\mp}(\overrightarrow{k})$ (27) are called the
spin states of the doublet. Nevertheless, they are not the
eigenvectors of the operator $s_{z}=s^{3}$ from
$\overrightarrow{s}$ (18). They are the eigenvectors of the
operator $s_{z}^{\mathrm{Dirac}}\equiv s^{3
\,\mathrm{Dirac}}\subset\overrightarrow{s}^{\mathrm{Dirac}}$ from
nonlocal spin operator in the PD-representation for the spinor field
$\psi$, which is given by

\begin{equation}\label{eq27}
\overrightarrow{s}^{\mathrm{Dirac}}=V^{-1}\overrightarrow{s}V=\overrightarrow{s}-\frac{i[\overrightarrow{\gamma}\times\nabla]}{2\widehat{\omega}}+\frac{\nabla\times[\overrightarrow{s}\times\nabla]}{\widehat{\omega}(\widehat{\omega}+m)}
\end{equation}

\noindent and which is obtained from the $\overrightarrow{s}$
(18) by the FW-transformation $V$ [18] (the prime operator $\overrightarrow{s}^{\mathrm{Dirac}}$ (27) coincide with the corresponding spin $-i\overrightarrow{s}^{\mathrm{FW}}$, where $\overrightarrow{s}^{\mathrm{FW}}$ is given in the table 1 in [18]). Indeed, it is easy to see
that the following equations are valid:

\begin{equation}\label{eq28}
s_{z}v_{\mathrm{1}}^{\mp}(\overrightarrow{k})\neq \frac{1}{2}v_{\mathrm{1}}^{\mp}(\overrightarrow{k}), \quad
s_{z}v_{\mathrm{2}}^{\mp}(\overrightarrow{k})\neq -\frac{1}{2}v_{\mathrm{2}}^{\mp}(\overrightarrow{k});\quad
s_{z}^{\mathrm{Dirac}}v_{\mathrm{1}}^{\mp}(\overrightarrow{k})= \frac{1}{2}v_{\mathrm{1}}^{\mp}(\overrightarrow{k}), \quad
s_{z}^{\mathrm{Dirac}}v_{\mathrm{2}}^{\mp}(\overrightarrow{k})= -\frac{1}{2}v_{\mathrm{2}}^{\mp}(\overrightarrow{k}),
\end{equation}

\noindent where $s_{z}=s^{3}(18),\,s_{z}^{\mathrm{Dirac}}=s^{\mathrm{Dirac}\,3}(27)$ (the assertion (28) is our small addition to the consideration [18]). Moreover, operator $\overrightarrow{s}$ (18) does not
commute with both the prime Dirac operator from equation (1) and
the operator $\partial_{0}-\widetilde{p}_{0}$ (or $H$) from
(14). Therefore, for the spinor field $\psi$ in the PD-representation
the operator $\overrightarrow{s}$ (18) can not be interpreted as
the quantum-mechanical spin operator for the spin
$\frac{1}{2}$-doublet of particles and does not generate the spin
conservation law (even in the absence of interaction), which contradicts the experiment. Furthermore, as is known
from [18], the operator
$\overrightarrow{x}=(x^{\ell})\in\mathrm{R}^{3}\subset\mathrm{M}(1,3)$
in the PD-representation for the field $\psi$ also cannot be
interpreted as the quantum-mechanical operator of 3-coordinate for
the spin $\frac{1}{2}$-doublet of particles. Nevertheless,
operators $\overrightarrow{s}$ (18) and $\overrightarrow{x}$ are
the important structure operators (which possess the physically
meaningful quantum-mechanical spectra) for the construction of
generators (4), (19) and corresponding local and induced
$\mathcal{P}^{\mathrm{F}}$-representations.

As shown in [18] the above shortcomings follow from the non-diagonality of the Hamiltonian
$H=i\widetilde{p}_{0}$ in (14). Therefore, in the PD-representation
of the field $\psi$, the particle and antiparticle states are
mixed. The progressive way of moving forward was suggested in [18]. The
FW-representation of the spinor field theory is free from the above shortcomings, e. g. the coordinate
$\overrightarrow{x}=(x^{\ell})$ of the FW-spinor $\phi$ and spin
$\overrightarrow{s}$ (18) in this representation have direct
physical meaning of corresponding quantum-mechanical observables
of spin-$\frac{1}{2}$ doublet of particles. It is important that
just the components of this spin $\overrightarrow{s}$ are the
elements of the CD-algebra in the PD-representation. The above arguments force us to start from the FW-representation of the
spinor field (not from the PD-representation) and to consider the
CD-algebra in the PD-representation just inside the
FW-representation of the spinor field.

Our search for \textit{the hidden symmetries} of the Dirac
equation is based on the method of \textit{extension} of the
CD-algebra as \textit{the real one} (on the basis of including in
it the orts $i$ and $C$). It appears to be possible just owing
to considering the $\mathcal{P}$-generators in terms of the primary
operators.

\section{The Foldy -- Wouthuysen representation and the extended real Clifford - Dirac algebra}

In order to derive the assertions mentioned in introduction, we
essentially use two constructive ideas.

(i) The above physical arguments (see Sec. 2) cause our start
not from the standard Dirac equation but from its FW
representation [18]

\begin{equation}
\label{eq29} (i\partial _0 - \gamma_{0}\widehat{\omega } )\phi (x)
= 0,\mbox{ }\phi=V\psi,
\end{equation}

\begin{equation}
\label{eq30}
V=\frac{-i\overrightarrow{\gamma }\cdot \nabla+\widehat{\omega }+m%
}{\sqrt{2\widehat{\omega }(\widehat{\omega }+m)}} \mbox{, }V^{ - 1} =
V(\nabla \to -
\nabla),\mbox{ }\widehat{\omega} \equiv \sqrt { -
\Delta + m^2}, \mbox{
} \nabla\equiv(\partial_{\ell}). \\
\end{equation}

In the space (12), the operator $V$ is unitary. The general
solution of the FW-equation (29) (the Dirac equation in the
FW-representation) in the quantum-mechanical rigged Hilbert space
(12) is given by

\begin{equation}\label{eq31}
\phi(x)=V\psi(x)=\frac{1}{(2\pi)^{\frac{3}{2}}}\int
d^{3}k[a^{\mathrm{r}}(\overrightarrow{k})\mathrm{D}_{\mathrm{r}}^{-}e^{-ikx}+
b^{*\mathrm{r}}(\overrightarrow{k})\mathrm{D}_{\mathrm{r}}^{+}e^{ikx}],\quad
\mathrm{r}=1,2,
\end{equation}

\noindent where the 4-dimensional Cartezian basis vectors in the
space $\mathrm{C}^{4}$ have the form

\begin{equation}\label{eq32}
\mathrm{D}_{\mathrm{r}}^{-}\equiv Vv_{\mathrm{r}}^{-}(\overrightarrow{k})=\left| {{\begin{array}{*{20}c}
 \mathrm{d}_{\mathrm{r}}\\
 0\\
\end{array} }} \right|,\quad \mathrm{D}_{\mathrm{r}}^{+}\equiv Vv_{\mathrm{r}}^{+}(\overrightarrow{k})=\left| {{\begin{array}{*{20}c}
 0\\
 \mathrm{d}_{\mathrm{r}}\\
\end{array} }} \right|,
\end{equation}

\noindent and amplitudes
$a^{\mathrm{r}}(\overrightarrow{k}),\,b^{\mathrm{r}}(\overrightarrow{k})$
as well as two-component vectors $\mathrm{d}_{\mathrm{r}}$ are
the same as in (25), (26). Contrary to the basis (26) the vectors
(32) are the eigenvectors of the quantum-mechanical spin operator
$s_{z}\in s^{3}$ from $\overrightarrow{s}$ (18). The
fundamental solutions $e^{\mp ikx}\mathrm{D}_{\mathrm{r}}^{\mp}$
of equation (29) are the relativistic quantummechanical de
Broglie waves (with the determined values of the spin $s_{z}$
from (18)) for the spin-$\frac{1}{2}$ doublet of particles.

Note that both the set of fundamental solutions $\{\Psi_{\vec{k}\mathrm{r}}^{\mp}\}$ of equation (1)=(14) and $\{\Phi_{\vec{k}\mathrm{r}}^{\mp}\}$ of the equation (29),

\begin{equation}\label{eq33}
\Psi_{\vec{k}\mathrm{r}}^{\mp}(x)\equiv \frac{1}{(2\pi)^{3/2}}e^{\mp ikx}v_{\mathrm{r}}^{\mp}(\overrightarrow{k}),\quad \Phi_{\vec{k}\mathrm{r}}^{\mp}(x)\equiv \frac{1}{(2\pi)^{3/2}}e^{\mp ikx}\mathrm{D}_{\mathrm{r}}^{\mp},
\end{equation}

\noindent are the common eigenvectors of the \textit{complete set
of mutually commute independent operators} in $\mathrm{S^{3,4}}$,
namely $(\overrightarrow{p}^{\mathrm{stand}}=(-i\partial_{\ell}),
\, s_{z}^{\mathrm{Dirac}})$ for the set
$\{\Psi_{\vec{k}\mathrm{r}}^{\mp}\}$ and $(\overrightarrow{p}, \,
s_{z})$ for the set $\{\Phi_{\vec{k}\mathrm{r}}^{\mp}\}$ (of
course, they satisfy one and the same orthonormality and
completeness conditions). These facts present the evident proof
of the adequate physical meaning of the spin operators
$\overrightarrow{s}^{\mathrm{Dirac}}$ (27) and
$\overrightarrow{s}$ (18): the first is the true spin for the
spinor field $\psi$ in PD-representation and the second is that for
the spinor field $\phi$ in FW-representation.

Moreover, now the $\mathcal{P}^{\mathrm{F}}$-generators of the unitary (induced) $\mathcal{P}^{\mathrm{F}}$-representation for the field $\phi$ (obtained from the operators (19) with the help of $V$-transformation (30)) are the functions of 11 directly experimentally observed operators $(\overrightarrow{x}=(x^{\ell}),\,
\overrightarrow{p}^{\mathrm{stand}}=(-i\partial_{\ell}), \,
\overrightarrow{s}^{\mathrm{stand}}=(is_{\ell n }),\,
\,\gamma^{0},\,m)$ or prime operators $(\overrightarrow{x}=(x^{\ell}),\,\nabla = (\partial_{\ell}),\,\overrightarrow{s}=(s^{\ell})(18),\,\gamma^{0},\,m)$:

\begin{equation}
\label{eq34}
 \check{p}_0 = - i\gamma
_{0}\widehat{\omega},\mbox{ }\check{p}_n=\widetilde{p}_{n}=\partial _n,\mbox{ }\check{j}_{ln}=\widetilde{j}_{ln} = x_l
\partial _n -
x_n \partial _l + {s}_{ln}, \,
\check{j}_{0k}=x_0\partial _k+i\gamma_{0}\{x_k\widehat{\omega }+\frac{\partial _k}{2\widehat{\omega }}
+\frac{(\overrightarrow{s}\times \overrightarrow{\partial })_k}{\widehat{\omega }+m}\}\\
\end{equation}

\noindent (the standard form of the primary generators (34) see e.g. in the formulae (D-64)--(D-67) in [24]). Now the $\mathcal{P}^{\mathrm{F}}$-representation for the field $\phi$,

\begin{equation}
\label{eq35}(a, \, \omega)\in \mathcal{P} \rightarrow
\check{\mathrm{U}}(a, \, \omega) = \exp
(a^{\mu}\check{p}_{\mu}+\frac{1}{2} \omega^{\mu\nu}
\check{j}_{\mu\nu}),
\end{equation}

\noindent (the $\mathcal{P}^{\mathrm{F}}$-group of invariance of
the equation (29)) have no of the above mentioned shortcomings
inherent for the local (3) and induced (20)
$\mathcal{P}^{\mathrm{F}}$-representations. The Casimir operators for the generators (34)
are the same as in (22), (23). The theory of the spinor field
$\phi$ in the FW-representation, which is recalled here, can be evidently generalized for the fields of any spin
$\overrightarrow{s}$ (where $(s^{\ell})\equiv \overrightarrow{s}$
are the arbitrary generators of an SU(2)-algebra representation).

Not a matter of special separate role of the time variable $t=x^{0}$,
the theory of the spinor field $\phi$ in the FW-representation is the
relativistic invariant (namely, $\mathcal{P}$-invariant) in the
sense of two necessary aspects. (i) The dynamical aspect: the
unitary $\mathcal{P}^{\mathrm{F}}$-representation (35) is the
group of invariance of the equation (29) (the main conservation
laws for the spinor field $\phi$ in the FW-representation are the
consequences of this assertion). (ii) The kinematical aspect: if
the solution (31) with arbitrary-fixed amplitudes
$a^{\mathrm{r}}(\overrightarrow{k}),\,b^{\mathrm{r}}(\overrightarrow{k})$
is a given state of the spinor field $\phi$ in the arbitrary-fixed inertial frame of references (IFR)
$\Sigma$, then for the observer in the $(a, \,
\omega)$-transformed IFR $\Sigma^{\prime}$ the solution
$\phi^{\prime}(x)=\check{\mathrm{U}}(a, \, \omega)\phi(x), \,
x\in\mathrm{M}(1,3)$, is the same state of the
spinor field.

Some other details of physical arguments for our start from the
FW-representation are as follows. Just the components of
the field $\phi$ of the FW equation (29) coincide with the
quantum-mechanical wave functions of the particle doublet.
Furthermore, it is the FW-representation, where the operators of
directly experimentally observed quantities of this doublet are
the corresponding direct sums of the quantum-mechanical observables
of single particles, which form the doublet. Therefore, it is in the
canonical (i.e. FW) representation where these operators have the
status of true observables of the particle doublet. In
particular, the 16-dimensional Clifford -- Dirac (CD) algebra
(see formulae (36) below) generated by the
$\gamma^{\mu}$-matrices (17) contains the generators of the SU(2)
group, which commute with the Hamiltonian $\gamma ^0\hat{\omega}$
of the FW equation (29). The last assertion means that the spin
$\overrightarrow{s}$ (18) of the free particle doublet  is
conserved. Therefore, the operator $\overrightarrow{s}$ (18)
given in the set $\{\phi\}$ of solutions (31) of equation (29) has the
status of true spin. This is the reason to consider the
prime CD algebra in the PD representation (17) as an algebra in
the set $\{\phi\}\subset (12)$. In order to avoid
misunderstanding, note that in our consideration the
$\gamma^{\mu}$-matrices in the FW representation have the form
(17), not $V\gamma^{\mu}V^{-1}$. Of course, $V^{-1}(i\partial _0 -
\gamma_{0}\widehat{\omega })V={i\gamma ^\mu \partial _\mu - m}$
with the $\gamma^{\mu}$-matrices (17).

Finally, one have a simple technical reason to prefer the
FW-representation. In the FW-representation the operator of equation (29) contains only one diagonal
$\gamma^{0}$-matrix instead of four $\gamma^{\mu}$-matrices in
equation (1)=(14). Hence, equation (29) and the
FW-representation are much more convenient in searching for the
matrix symmetries (in comparison with the prime equation (1)=(14) and
the PD-representation for the spinor field). Note that such method and these
advantages of the FW-representation are well-known since the appearance of
publications [18, 24] (see e.g. the table 1 in [18], where the important for physics symmetry operator $\overrightarrow{s}^{\mathrm{Dirac}}$ (27) was found, or the generators (126)--(129) in [24]). We recall them only for our purposes.

(ii) Here (see the formula (38) below) we introduce into
consideration a 64-dimensional extended real Clifford -- Dirac
algebra (ERCD algebra) as the algebra of $4\times4$ \textit{pure
matrix operators} in the set $\{\phi\}\subset(12)$. We essentially
apply it here as a constructive mathematics for our purposes. For
the physical purposes, when the \textit{parameters} of the
relativistic groups are \textit{real}, it is sufficient to
consider the 16-dimensional standard CD algebra in the complex
space (12) as a \textit{real algebra}. It enables us, along with
the application of the imaginary unit "$i$" and the operator
$\hat{C}$ of complex conjugation in the set
$\{\phi\}\subset(12)$, to extend this algebra to the ERCD algebra
in the space (12). It is on the base of the ERCD algebra that we
are able to find the maximal pure matrix algebra (without the
space-time derivatives $\partial_{\mu}$) of invariance of the FW
equation and the corresponding algebra of invariance of the Dirac
equation with an arbitrary mass in the standard PD
representation. The ERCD algebra is the complete set of
operators, the part of which generates the Pauli -- Gursey --
Ibragimov (PGI) algebra. Let us recall that the 8-dimensional (for $m$=0) and
4-dimensional (for a nonzero mass) PGI
algebras of invariance of the Dirac equation are considered in details, for instance,
in [13,17] (these algebras were put into consideration in the original papers of W. Pauli, F. Gursey and N. Ibragimov, see, e. g.
[25,26]).

Finally, it is useful to choose 16 independent (ind) generators
-- orts of standard CD algebra as

\begin{equation}
\label{eq36}\left\{ {\mbox{indCD}} \right\} \equiv \left\{
\mbox{I},\mbox{ }\alpha_{\hat{\mu}\hat{\nu}} =
2{s}_{\hat{\mu}\hat{\nu}}:{s}_{\check{\mu}\check{\nu}} \equiv
\frac{1}{4}\left[ \gamma_{\check{\mu}},\gamma_{\check{\nu}}
\right],\mbox{ }{s}_{\check{\mu}5} = - {s}_{5\check{\mu}}\equiv
\frac{1}{2}\gamma_{\check{\mu}};\mbox{ }\gamma_{4}\equiv
\gamma_{0}\gamma_{1}\gamma_{2}\gamma_{3},\mbox{ }
\hat{\mu},\hat{\nu}=\overline{0,5},\mbox{
}{\check{\mu},\check{\nu} = \overline{0,4}}\right\},
\end{equation}

\noindent where
$\gamma_{4}\equiv\gamma_{0}\gamma_{1}\gamma_{2}\gamma_{3}=
i\gamma_{5}^\mathrm{stand}$ and the matrices
${s}_{\hat{\mu}\hat{\nu}}$ are the \textit{prime generators} of
the SO(1,5)$\supset$SO(1,3)$=\mbox{L}_ + ^ \uparrow$ group
(associated with the real parameters -- the angles
$\omega^{\hat{\mu}\hat{\nu}}$ of rotations in
$\hat{\mu}\hat{\nu}$ planes of the space
$\mbox{M(1,5)}\supset\mbox{M(1,3)}$), these operators satisfy the
following commutation relations

\begin{equation}
\label{eq37}\left[ {{s}_{\hat{\mu}\hat{\nu}},
{{s}_{\hat{\rho}\hat{\sigma}}}}\right] = -
{g}_{\hat{\mu}\hat{\rho}} {s}_{\hat{\nu}\hat{\sigma}} -
{g}_{\hat{\rho}\hat{\nu}} {s}_{\hat{\sigma}\hat{\mu}} -
{g}_{\hat{\nu}\hat{\sigma}} {s}_{\hat{\mu}\hat{\rho}} -
{g}_{\hat{\sigma}\hat{\mu}} {s}_{\hat{\rho}\hat{\nu}};\mbox{ }\hat{\mu},\hat{\nu}=\overline{0,5},\mbox{ }
(g_{\hat{\nu}}^{\hat{\mu}})=\mathrm{diag}(+1,-1,-1,-1,-1,-1).
\end{equation}

By complementing the orts (36) of the real CD algebra with the
operators $i$, $\hat{C}$ and with all possible
products of orts (36) and operators $i$ and $\hat{C}$, we define
the ERCD algebra as the linear manifold spanned on the orts

\begin{equation}
\label{eq38} \left\{ {\mbox{ERCD}} \right\} = \left\{
{\mbox{indCD, }i \cdot \mbox{indCD, } \hat{C} \cdot \mbox{indCD,
}i\hat{C} \cdot \mbox{indCD}} \right\}.
\end{equation}

Thus, the ERCD algebra generators are the compositions of the
standard CD algebra generators (36) and the generators of the
PGI algebra [25, 26], i. e. it is the
maximal set of independent matrices, which can be constructed from
the elements $i$, $\hat{C}$, and (36).

All the physically meaningful symmetries of the FW and Dirac
equations put into consideration below are
constructed using the elements of the ERCD algebra.

\section{Maximal pure matrix algebra of invariance of the Foldy -- Wouthuysen equation}

Consider the 32-dimensional subalgebra $\mbox{A}_{32} =
\mbox{SO(6)} \oplus \hat{\varepsilon} \cdot \mbox{SO(6)} \oplus
\hat{\varepsilon}$ of the ERCD
algebra. It is easy to see that the part of ERCD algebra
generators, namely the operators

\begin{equation}
\label{eq39} s_{AB}=\frac
14[\gamma_{A},\gamma_{B}]=-s_{BA},\quad\mbox{
}A,B=\overline{1,6},\quad\mbox{ }\gamma_{5}\equiv
\gamma_{1}\gamma_{3}\hat{C},\quad\mbox{ }\gamma_{6}\equiv
i\gamma_{1}\gamma_{3}\hat{C},
\end{equation}

\noindent (here $A,B=\overline{1,6}$ includes
$\check{a}\check{b}=\overline{1,4}$ from (36)) and
$\hat{\varepsilon} = i\gamma_{0}$, satisfy the commutation
relations

\begin{equation}
\label{eq40} \left[ {s}_{AB} ,{s}_{CD} \right] = \delta_{AC}
{s}_{BD} + \delta_{CB} {s}_{DA} + \delta_{BD} {s}_{AC} +
\delta_{DA} {s}_{CB}, \mbox{ }\left[ {s}_{AB} , \hat{\varepsilon}
\right] = 0; \mbox{ } A, B, C, D = \overline{1,6}.
\end{equation}

\noindent Therefore, operators (39) generate a representation of the $\mbox{SO(6)}\supset\mbox{SO(3)}$
group of space rotations in the space
$\mathrm{R}^{6}\subset\mathrm{M}(1,6)\supset\mathrm{M}(1,3)$. On
this basis, together with the additional $\mbox{SO(6)}$ Casimir
operator $\hat{\varepsilon}\equiv i\gamma_{0} = -
\gamma_{1}\gamma_{2}\gamma_{3}\gamma_{4}\gamma_{5}\gamma_{6}$, we
define the $\mbox{SO(6)} \oplus \hat{\varepsilon}$ algebra and,
finally, the $\mbox{A}_{32} = \mbox{SO(6)} \oplus
\hat{\varepsilon} \cdot \mbox{SO(6)} \oplus \hat{\varepsilon}$
algebra. The commutation relations for the $\hat{\varepsilon}
\cdot \mbox{SO(6)}$ generators ${\tilde{s}_{AB}} =
\hat{\varepsilon} s_{AB}$ differ from (40) by the common factor
$\hat{\varepsilon} = i\gamma_{0}$. \textbf{\textit{The
32-dimensional subalgebra $\mbox{A}_{32} = \mbox{SO(6)} \oplus
\hat{\varepsilon} \cdot \mbox{SO(6)} \oplus \hat{\varepsilon}$ of the
ERCD algebra in the rigged Hilbert space (12) is the maximal pure matrix algebra (without the space-time
derivatives $\partial_{\mu}$) of invariance of the
FW-equation (29)}}.

The \textbf{\textit{proof}} of this assertion is carried out by
straightforward calculations of (i) the corresponding commutation
relations (40) for elements of this algebra, (ii) the commutators
between the elements of A$_{32}$ and the operator $(i\partial _0 -
\gamma_{0}\hat{\omega})$ of the FW equation (29). The maximality
of $\mbox{A}_{32} $ as the algebra of invariance of the equation
(29) is the consequence of the maximality of the dim(ERCD)=64 (in the
class of pure matrix operators).

Note that antihermitian matrices
$\{\gamma_{A}:\,\gamma_{1},\,\gamma_{2},\,\gamma_{3},\,\gamma_{4}\equiv\gamma_{0}\gamma_{1}\gamma_{2}\gamma_{3},\,\gamma_{5}\equiv
\gamma_{1}\gamma_{3}\widehat{C},\,\gamma_{6}\equiv
i\gamma_{1}\gamma_{3}\widehat{C},\,\gamma_{7}\equiv
i\gamma_{0}\}$ satisfy the commutation relations $
\gamma_{A}\gamma_{B} + \gamma_{B}\gamma_{A} = -2\delta_{AB},\,
A,B=\overline{1,7}$.

The explicit form of the elements of corresponding algebra
$\mbox{A}_{32}$ of invariance of the Dirac equation in the
PD-representation is found from the elements (39) and
$\hat{\varepsilon}$ with the help of the FW-transformation (30):
$V^{-1}(\mbox{A}_{32},\,\hat{\varepsilon})V$. In
PD-representation this algebra of invariance (of the prime Dirac equation (1)) is given by the nonlocal
operators.

\section{Spin 1 Lorentz-symmetries of the Foldy -- Wouthuysen and Dirac equations}

The FW-equation (29) is invariant with respect to the two
different spin 1 representations of the Lorentz group
$\mathcal{L}$ (below ${s}_{\mu \nu }^\mathrm{V}$ are the
generators of the irreducible vector (1/2,1/2) representation and ${s}_{\mu \nu
}^\mathrm{TS}$ are the generators of the reducible tensor-scalar
$(1,0) \oplus (0,0)$ representation of the SO(1,3)$=\mbox{L}_ +
^ \uparrow $ algebra). The explicit forms of the corresponding
pure matrix operators are given by

\begin{equation}
\label{eq41} {s}_{\mu \nu}^\mathrm{TS} = \{{s}_{0k}^\mathrm{TS} =
{s}_{0k}^\mathrm{I} +
{s}_{0k}^\mathrm{II},\quad{s}_{mn}^\mathrm{TS} =
{s}_{mn}^\mathrm{I} + {s}_{mn}^\mathrm{II}\},\quad\mbox{ }
{s}_{\mu \nu }^\mathrm{V} = \{ {s}_{0k}^\mathrm{V} = -
{s}_{0k}^\mathrm{I} +
{s}_{0k}^\mathrm{II},\quad{s}_{mn}^\mathrm{V} =
{s}_{mn}^\mathrm{TS} \},
\end{equation}

\noindent where $\mathrm{s}_{\mu\nu}^{\mathrm{I,II}}$ are the following elements of $\mathrm{A}_{32}$ algebra:

\begin{equation}
\label{eq42} {s}_{\mu \nu }^\mathrm{I} = \{{s}_{0k}^\mathrm{I} =
\frac{i}{2}\gamma_{k}\gamma_{4},\quad\mbox{ }{s}_{mk}^\mathrm{I} =
\frac{1}{4}[\gamma_{m},\gamma_{k}]\},\quad\mbox{ }\gamma_{4}
\equiv \gamma_{0}\gamma_{1}\gamma_{2}\gamma_{3},\quad\mbox{ }(k,m
= \overline {1,3}),
\end{equation}

\begin{equation}
\label{eq43} {s}_{\mu \nu }^\mathrm{II} = \{{s}_{01}^\mathrm{II} =
\frac{i}{2}\gamma_{2}\hat{C},\quad{s}_{02}^\mathrm{II} = -
\frac{1}{2}\gamma_{2}\hat{C},\quad{s}_{03}^\mathrm{II} = -
\frac{1}{2}\gamma_{0},\quad{s}_{12}^\mathrm{II} =
\frac{i}{2},\quad{s}_{31}^\mathrm{II} =
\frac{i}{2}\gamma_{0}\gamma_{2}\hat{C},\quad{s}_{23}^\mathrm{II} =
\frac{1}{2}\gamma_{0}\gamma_{2}\hat{C}\}.
\end{equation}

\noindent The sets $\mathrm{s}_{\mu\nu}^{\mathrm{I}}$ (42) and $\mathrm{s}_{\mu\nu}^{\mathrm{II}}$ (43) determine in $\mathrm{A}_{32}$ the generators of two different versions of $(1 / 2,0) \oplus (0,1 / 2)$ representation of the
$\mathcal{L}$-algebra. The validity of these assertions are evident after the
transition ${s}_{\mu \nu }^\mathrm{Bose} = W{s}_{\mu \nu
}^\mathrm{V,TS}W^{ - 1}$ in the Bose-representation of the
$\gamma $-matrices (${\gamma}^{\mu \,\mathrm{Bose}} =
W{\gamma}^{\mu}W^{ - 1}$), where the operator $W$ is given by

\begin{equation}
\label{eq44} W = \frac{1}{\sqrt 2}\left|
\begin{array}{cccc}
 0 & -1 & 0 & \hat{C}\\
 0 & i & 0 & i\hat{C}\\
-1 & 0 & \hat{C} & 0\\
-1 & 0 &  -\hat{C} & 0\\
\end{array} \right|,
\mbox{ } W^{-1}= \frac{1}{\sqrt 2}\left|
\begin{array}{cccc}
0 & 0 & -1 & -1\\
-1 & -i & 0 & 0\\
0 & 0 & \hat{C} & -\hat{C}\\
\hat{C} & i\hat{C} & 0 & 0\\
\end{array}\right|,\mbox{ } WW^{-1}=W^{-1}W=1.
\end{equation}

\noindent In such Bose-representation, the ${\gamma}^{\mu \,
\mathrm{Bose}}$-matrices contain not only the operator $i$, but
also the operator $\hat{C}$ of complex conjugation. Nevertheless,
the Lorentz spin matrices ${s}_{\mu \nu }^\mathrm{Bose} =
W{s}_{\mu \nu }^\mathrm{V,TS}W^{ - 1}$ in the Bose-representation
do not contain the operator $\hat{C}$ and take the explicit forms
known well for the matrix representations $(1,0) \otimes (0,0)$
and (1/2,1/2) of the group $\mathcal{L}\sim$SO(1,3):

$${s}_{12}^\mathrm{V,TS\,Bose}=\left|
\begin{array}{cccc}
 0 & -1 & 0 & 0\\
 1 & 0 & 0 & 0\\
0 & 0 & 0 & 0\\
0 & 0 & 0 & 0\\
\end{array} \right|,\quad{s}_{31}^\mathrm{V,TS\,Bose}=\left|
\begin{array}{cccc}
 0 & 0 & 1 & 0\\
 0 & 0 & 0 & 0\\
-1 & 0 & 0 & 0\\
0 & 0 & 0 & 0\\
\end{array} \right|,\quad{s}_{23}^\mathrm{V,TS\,Bose}=\left|
\begin{array}{cccc}
 0 & 0 & 0 & 0\\
 0 & 0 & -1 & 0\\
0 & 1 & 0 & 0\\
0 & 0 & 0 & 0\\
\end{array} \right|,$$

\begin{equation}
\label{eq45} {s}_{01}^\mathrm{TS\,Bose}=\left|
\begin{array}{cccc}
 0 & 0 & 0 & 0\\
 0 & 0 & -i & 0\\
0 & i & 0 & 0\\
0 & 0 & 0 & 0\\
\end{array} \right|,\quad{s}_{02}^\mathrm{TS\,Bose}=\left|
\begin{array}{cccc}
 0 & 0 & i & 0\\
 0 & 0 & 0 & 0\\
-i & 0 & 0 & 0\\
0 & 0 & 0 & 0\\
\end{array} \right|,\quad{s}_{03}^\mathrm{TS\,Bose}=\left|
\begin{array}{cccc}
 0 & -i & 0 & 0\\
 i & 0 & 0 & 0\\
0 & 0 & 0 & 0\\
0 & 0 & 0 & 0\\
\end{array} \right|,
\end{equation}

$${s}_{01}^\mathrm{V\,Bose}=\left|
\begin{array}{cccc}
 0 & 0 & 0 & -1\\
 0 & 0 & 0 & 0\\
0 & 0 & 0 & 0\\
-1 & 0 & 0 & 0\\
\end{array} \right|,\quad{s}_{02}^\mathrm{V\,Bose}=\left|
\begin{array}{cccc}
 0 & 0 & 0 & 0\\
 0 & 0 & 0 & -1\\
0 & 0 & 0 & 0\\
0 & -1 & 0 & 0\\
\end{array} \right|,\quad{s}_{03}^\mathrm{V\,Bose}=\left|
\begin{array}{cccc}
 0 & 0 & 0 & 0\\
 0 & 0 & 0 & 0\\
0 & 0 & 0 & -1\\
0 & 0 & -1 & 0\\
\end{array} \right|.$$

We call the transition $\phi\rightarrow W\phi$ with $W$ (44),
which links the given fermionic and bosonic multipletes, a new
natural form of the supersymmetry transformation. The corresponding symmetries of the Dirac equation are obtained with the help of FW-transformation (30).

\section{Spin 1 Poincare-symmetries of the Foldy -- Wouthuysen equation}

\textbf{\textit{The FW equation (29) is invariant}} not only with respect to the
well-known standard spin 1/2 $\mathcal{P}^{\mathrm{F}}$-representation (35), but \textbf{\textit{also with respect to the canonical-type spin 1 representation (Bose representation) of the
Poincare group}} $\mathcal{P}$, i.e. with respect to the unitary
(in the set $\{\phi \}$ of solutions of equation (29))
$\mathcal{P}^{\mathrm{B}}$-representation, which is determined by the primary
generators

\begin{equation}
\label{eq46}
 \widehat{p}_0=\check{p}_0 = - i\gamma
_{0}\widehat{\omega},\mbox{ }\widehat{p}_n=\check{p}_n=\partial
_n,\mbox{ }\widehat{j}_{ln} = x_l
\partial _n -
x_n \partial _l + {s}_{ln}^\mathrm{I} + {s}_{ln}^\mathrm{II},\\ \,
\widehat{j}_{0k}=x_0\partial _k+i\gamma_{0}\{x_k\widehat{\omega
}+\frac{\partial _k}{2\widehat{\omega }}
+\frac{[(\overrightarrow{s}^\mathrm{I}+\overrightarrow{s}
^\mathrm{II})\times \overrightarrow{\partial }]_k}{\widehat{\omega
}+m}\}\\,
\end{equation}

\noindent where $\widehat{\omega}$ is given in (30),
${s}_{ln}^\mathrm{I}$ and ${s}_{ln}^\mathrm{II}$ are given in
(42), (43), respectively, and $\vec {{s}}^\mathrm{I,II} =
({s}_{23} ,{s}_{31} ,{s}_{12})^\mathrm{I,II}$.

The \textbf{\textit{proof}} is performed by the straightforward
calculations of (i) the corresponding $\mathcal{P}$-commutators (2)
between the generators (46), (ii) the commutators between
generators (46) and operator $(i\partial _0 -
\gamma_{0}\hat{\omega})$, (iii) the Casimir operators of the
Poincare group for the generators (46). Acoording to the Bargman -- Wigner classification
of the $\mathcal{P}$-covariant fields, just these facts
(especially (iii)) visualize the hidden Bose essence of the
$\mathcal{P}^{\mathrm{B}}$-representation, generated by the operators (46).
For the $\mathcal{P}^{\mathrm{B}}$-representation the explicit form of main Casimir operators is following

\begin{equation}
\label{eq47}\widehat{p}^{\mu}\widehat{p}_\mu = m^{2}, \quad
\mathcal{W}^\mathrm{B}=w^{\mu}w_\mu =
m^{2}(\overrightarrow{s}^\mathrm{TS})^{2},
\end{equation}

\noindent (compare with (22), (23)), where $w^{\mu}\equiv
\frac{1}{2}\varepsilon^{\mu\nu\rho\sigma}\widehat{p}_{\rho}\widehat{j}_{\nu\sigma}$ and, after diagonalization carried out
with the help of operator $W$ (44),

\begin{equation}
\label{eq48}(\overrightarrow{s}^\mathrm{TS})^{2} =
(\overrightarrow{s}^\mathrm{TS \, Bose})^{2} = -1(1+1) \left|
\begin{array}{cccc}
 \mbox{I
}_{3} & 0\\
 0 & 0\\
\end{array} \right|, \quad \mathrm{I}_{3}\equiv\left|
\begin{array}{cccc}
 1 & 0 & 0 &\\
 0 & 1 & 0 &\\
0 & 0 & 1 &\\
\end{array} \right|.
\end{equation}

\section{Spin 1 symmetries of the Dirac equation with nonzero mass}

It is easy to see that \textbf{\textit{the prime Dirac equation (\ref{eq1}) has all
above mentioned spin 1 symmetries of the FW equation}}. The
corresponding explicit forms of the generators $q^\mathrm{PD}$ in
the manifold $\{\psi \}$ are obtained from the corresponding
formulae (39), (41) -- (43), (46) for the FW generators
$q^\mathrm{FW}$ with the help of the FW operator $V$ (30):
$q^\mathrm{PD} = V^{-1}q^\mathrm{FW} V$. As a meaningful example,
we present here the explicit form for \textbf{\textit{the spin 1 generators of
$\mathcal{P}^{\mathrm{B}}$-symmetries of the Dirac equation}}

\begin{equation}
\label{eq49} \widehat{p}_{0}^{\mathrm{PD}}=\widetilde{p}_{0}=-iH,\mbox{ }\widehat{p}_{k}^{\mathrm{PD}} =
\partial_{k} ,\mbox{ }\widehat{j}_{kl}^{\mathrm{PD}} = x_k
\partial _l - x_l
\partial _k + {s}_{kl} + \hat{s}
_{kl},\mbox{ }\widehat{j}_{0k}^{\mathrm{PD}} = x_0 \partial _k -
x_k \widetilde{p}_{0} + {s}_{0k} + \frac{\varepsilon _{k\ell
n}\hat{s}_{0\ell} \partial _n}{\widehat{\omega} + m},
\end{equation}

\noindent where $\varepsilon _{kln} $ is the Levi-Chivitta tensor,
and the operators ${s}_{\mu \nu }$, $\hat{s}_{\mu \nu } = V^{ - 1}
{s}_{\mu \nu }^\mathrm{II}V$ have the form

$${s}_{\mu \nu } = \frac{1}{4}\left[ {\gamma _\mu ,\gamma _\nu }
\right], \quad \mu,\nu = \overline{0,3},$$

\begin{equation}
\label{eq50} \hat{s}_{\mu\nu}=\{\hat{s}_{01}=
\frac{1}{2}i\gamma_{2}\hat{C},\mbox{ }\hat{s}_{02}=-
\frac{1}{2}\gamma_{2}\hat{C} ,\mbox{ }\hat{s}_{03}=-\gamma
^0\frac{\overrightarrow{\gamma}\cdot
\overrightarrow{p}+m}{2\widehat{\omega}}=-\frac{H}{2\widehat{\omega}},
\end{equation}

$$\hat{s}_{12} = \frac{i}{2},\mbox{ }\hat{s}_{31} = \frac{iH
}{2\widehat{\omega}}\gamma_{2}\hat{C} ,\mbox{ }\hat{s}_{23} =
\frac{H}{2\widehat{\omega}}\gamma_{2}\hat{C}\}$$

\noindent (the part of the Lorentz spin operators from (50) is not
pure matrix because they depends on the pseudodifferential operator
$\widehat {\omega } \equiv \sqrt { - \Delta + m^2}$ well-defined
in the space $\mathrm S^{3,4}$). Of course, the Casimir operators for the $\mathcal{P}^{\mathrm{B}}$-generators (49)
have the same final form (47), (48) as for the generators (46). The
Dirac equation in the Bose-representation of the $\gamma
$-matrices (${\gamma}^{\mu \,\mathrm{Bose}} = W{\gamma}^{\mu}W^{
- 1}$) is the Maxwell-type equation for a massive tensor-scalar
field.

Note that the generators (46) without the additional terms
${s}_{ln}^\mathrm{II}$ (43) and $\vec {{s}}^\mathrm{II} =
({s}_{23} ,{s}_{31} ,{s}_{12} )^\mathrm{II}$ directly coincide
with the well-known generators (34) of standard Fermi (spin 1/2)
$\mathcal{P}^{\mathrm{F}}$-symmetries of the FW-equation (similar situation occurs for the generators (49) taken without
the terms including the operators $\hat{s} _{\mu \nu } $ from
(50) -- they coincide with the operators (19) of the induced $\mathcal{P}^{\mathrm{F}}$-representation (20)). These well-known
forms determine the Fermi-case while operators
suggested here are related to the Bose interpretation of
equations (1), (29), which is found here also to be possible. The
only difference of our Fermi-case from the spin $1/2$ generators
in [18] is that we use the prime form of generators
related to the real parameters of the Poincare group.

\section{The case of zero mass, brief remix}

After the analysis of the arbitrary mass case presented
here, the specific character of the zero mass case [12--17]
becomes evident. For $m=0$ the analogs of the additional
operators $\hat{s}_{\mu\nu}$ (50) are the local pure matrix Lie
operators (see, e. g. formulae (12) in [13]). Therefore, for the
$m=0$ case, the corresponding $\mathcal{P}^{\mathrm{F}}$-generators are
much simpler than those in (49) and contain no nonlocal
terms. Moreover, for the $m=0$ case, the appealing to the
FW-representation and ERCD-algebra is not necessary. In [12--17]
all Bose symmetries of the massless Dirac equation were found in the
standard PD-representation of the spinor field $\psi$ on the
basis of the ordinary CD-algebra and well-known [25, 26] PGI operators. Thus, the search for the
additional bosonic symmetries of the massless Dirac equation is
technically much easier. Therefore, our results [12--17]
contain a lot of additional meaningful information in comparison
with the results presented here for the nonzero mass. In [12--17], the
full consideration of the Fermi -- Bose duality of the massless Dirac
equation is given. Furthermore, the Fermi -- Bose duality of the
Maxwell equations with the gradient-type sources, which are the Bose
partner of the massless Dirac equation, is investigated in
details. In our further publications we will be in position to extend
and generalize on the same level all our results presented
briefly here and to reproduce all the results of [12--17] for the
case of nonzero mass. For example, the unitary relationship
between the fermionic amplitudes
$a^{\mathrm{r}}(\overrightarrow{k}),\,b^{\mathrm{r}}(\overrightarrow{k})$
of the solution (25) and the bosonic amplitudes of the general
solution of the Maxwell-type equation for a massive tensor-scalar
field will be given (the analog of the unitary relationship (31)
in [15], or (40) in [16], from the case $m=0$). Hence, the next
steps of the analysis of the Fermi -- Bose duality of the Dirac equation
with arbitrary mass will be given.

\section{Brief conclusions}

The following four principal results have been proved and presented.

1. In the FW-representation for the Dirac equation the new
mathematical object -- the 64-dimensional Extended real Clifford
-- Dirac (ERCD) algebra -- is put into consideration. The ERCD
algebra is a pure matrix algebra, i. e. the algebra without any
derivatives from the space variables, transformations of reflections,
inversions etc.

2. The 32-dimensional subalgebra $\mbox{A}_{32} = \mbox{SO(6)}
\oplus i\gamma_{0} \cdot \mbox{SO(6)} \oplus i\gamma_{0}\; (i\gamma_{0}=\hat{\varepsilon})$ of the
ERCD algebra is proved to be new and the maximal pure matrix
algebra of invariance of the FW equation. Its image
$V^{-1}(\mbox{A}_{32})V$ is the algebra of invariance of the
standard Dirac equation in PD-representation.

3. It is shown that some subsets of generators from the
$\mbox{A}_{32}$ have the meaning of the well-known fermionic SL(2,C)-
and SU(2)-spins (see formulae (42), (43)), and other (41) have the meaning of new bosonic
SL(2,C)- and SU(2)-spins. Therefore, as the symmetry operators,
they have the useful physical interpretation.

4. On the basis of the above (see item 3) spins, operators
$\overrightarrow{x}$ and $\nabla$ (together with well-known
fermionic $\mathcal{P}^{\mathrm{F}}$-representation), the new
(hidden) bosonic $\mathcal{P}^{\mathrm{B}}$-representation of the
proper ortochronous Poincare group $\mathcal{P}$, as the group of
invariance of the Dirac and FW equations, has been found. Using the
Bargmann -- Wigner analysis this new $\mathcal{P}$-representation
has been proved to be the $\mathcal{P}^{\mathrm{B}}$--representation for
the Dirac field as the field of particles with the spin 1.

Therefore, here the assertion that "the Dirac equation can describe
the bosons, not only the fermions" is put into consideration as
an interesting new possibility. The details and consequences of
this assertion will be given in our further publication, where all
results of our papers [12--17] on the $m=0$ case will be
repeated for the general case of nonzero mass. Thus, we suggest
here a begining of a natural approach to the supersymmetry of the
Dirac equation, in which the Fermi and Bose superpartners are
linked by the transformation (44).

The possibilities of the ERCD algebra application are much more
extended than a few examples with the Dirac and FW equations
considered above. In general, the ERCD algebra can be applied to
any problem, in which the standard CD algebra can be used.
Therefore, the introduction of the ERCD algebra is a meaningful
independent result in the field of mathematical physics.

\end{document}